\documentclass[a4paper,12pt]{article}
\usepackage{epsfig}
\usepackage{amssymb}
\usepackage{amsmath}
\begin{document}
\thispagestyle{empty}

\newcommand{\etal}  {{\it{et al.}}}  
\def\Journal#1#2#3#4{{#1} {\bf #2}, #3 (#4)}
\def\PRD{Phys.\ Rev.\ D}
\def\NIMA{Nucl.\ Instrum.\ Methods A}
\def\PRL{Phys.\ Rev.\ Lett.\ }
\def\PLB{Phys.\ Lett.\ B}
\def\EPJ{Eur.\ Phys.\ J}
\def\IEEETNS{IEEE Trans.\ Nucl.\ Sci.\ }
\def\CPCD{Comput.\ Phys.\ Commun.\ }


\bigskip

{\Large\bf
\begin{center}
The scalar garlands in the boson star   
\end{center}
}
\vspace{0.1 cm}

\begin{center}
{ G.A.  Kozlov  }
\end{center}
\begin{center}
\noindent
 { Bogolyubov Laboratory of Theoretical Physics\\
 Joint Institute for Nuclear Research,\\
 Joliot Curie st., 6, Dubna, Moscow region, 141980 Russia  }
\end{center}
\vspace{0.1 cm}

 \begin{abstract}
 \noindent
 {
 In the paper, we introduce the formalism to examine the impact of the hidden scalar sector to the conformal and the electroweak symmetries breaking. The novel approach to the scalar boson star (BS) naturalness is considered. The BS is presented by the local scalar field containing the Higgs boson field and the garland-like scalar dilaton fields of the conformal field theory.
 We show that taking into account the repulsive self-interactions and the flatness degree in the dark scalar sector prevents instability of the BS related to the black hole formation.
 We study in details how electroweak symmetry breaking affects the hidden sector by breaking its conformal symmetry and generating a mass gap to avoid the infra-red divergence.
 We apply our formalism to determine the modification of the Higgs quartic coupling away from its standard model (SM) value within the influence of the hidden sector. We have estimated the rate of the deviation from the SM with production of  leptonic pairs due to decay of the dilaton and the dark photon. The latter is  the subject of the dilaton decay first. 
 The effect of new physics should be visible at the maximal energies of the LHC and the FCC.}


\end {abstract}




\bigskip

{\it Introduction. -}
Any phenomena in the Universe are governed by the principles of symmetry. Since the triumph of the Higgs discovery [1,2] which is based on the electroweak (EW) symmetry and its breaking [3,4], there are unexplained rates of both the Higgs mass and the vacuum expectation value (VEV) $ v$ where the latter is the measure of the mass scales in the standard model (SM) of elementary particles.  
The SM is widely regarded as an effective theory below the EW scale $\sim$ TeV, set by $v$. New physics (NP) at multi-TeV is currently actively explored at the LHC and in astroparticle experiments to find a hint for physics beyond the SM, though no any clear signals of NP have been found so far. 
Recent evidence, the theoretical and the phenomenological studies have emphasised the potentially crucial role played by evolving scalar fields, called the "dilaton" fields, in  understanding the cosmological processes in the early Universe. There is the possibility that various new vector bosons in hidden sector, e.g., dark photons (DP) dynamically related with dilatons, comprise the cosmological missing mass.
The scale invariance and the breaking of the invariance  are related with a new scalar field.
This means one may have in the spectrum the CP even scalar field that can be identified as the pseudo-Goldstone  boson associated with the spontaneous breaking of conformal symmetry at the scale $ f \geq v$. 
The model with the effective Lagrangian density (LD) should still be scale-invariant, and the effective scalar potential should preserve the pattern of the classical potential. We shall study the model containing both the Higgs boson and the dilaton, where all the scales are induced by spontaneous breakdown of scale invariance. The light dilaton itself decoupled from the SM sector can be associated with the dark matter (DM) or even with dynamical dark energy in the early Universe where the dynamical breaking of scale symmetry by the Higgs can explain a mechanism for the inflation [5].
Like in the Higgs case, the mass of the dilaton, is proportional to $f$ where the coefficient before $f$ is the parameter defined by the dynamics of the dilaton and indicates the rate of deviations from exact scale invariance [6-16].  If the SM is embedded in the conformal field theory (CFT), the dilaton is an important example that can allow to go beyond the SM. In some cases, the dilaton may have the mixing with the SM Higgs boson at energies below $f$ resulting with two physical scalar states each with production, e.g., by the colliders or due to an annihilation of spin $1/2$ DM  in case of the dilatons. In the paper [15], the authors introduced a new approach to the Higgs naturalness problem where the main experimental prediction is a dilaton in the 0.1 - 10 GeV range that mixes with the Higgs. The mixing angle is proportional to the ratio between the masses squared of the dilaton and the Higgs boson. As to the collider physics, there is the region of parameter space where the existence of a light dilaton is compatible with the LHC exclusion limits and where the hierarchy between the dilaton mass and $f$ is of the order $\sim O(1-10 \%)$. This represents the new physics that dedicated the collider searches for light new scalar states, e.g., the long-lived particles (LLP), in the LHC Run 3 that will be able to further explore and test. 
In the paper [17], it has been shown that in the range of the dilaton mass [60 - 160] GeV the LHC direct searches in the di-jets and the di-photon channels are dominant and a lower bound for $f$ is 3 TeV. The upper limit of $f$ is estimated in [18]: $f < $ 5.33 TeV if the dilaton is lighter than the top quark, or  $f < $ 4.87 TeV otherwise. 
The experimental and theoretical  results, together with several strong constraints in flavour physics and charged lepton physics indicate that there might be no NP with direct and sizeable couplings to SM sector up to energy $M_{NP}\sim 10^{5}$ TeV  [19] unless specific symmetries and breaking down are assumed or/and postulated.  

The determination of the Higgs VEV and the VEV of the dilaton are  still the open questions. The properties of the dilaton are similar to those of the SM Higgs boson. A distinctive difference between the dilaton and the Higgs is in their couplings to massless gauge bosons. At energies below $4\pi f$ the dilaton has the coupling with the trace of the SM  energy-momentum tensor where the quantum loop effects due to the conformal anomaly are included [20]. 
Here, the photons, the gluons, the DM hidden sector, e.g., dark photons, are contributed to the calculations. The $\beta$-functions for an electromagnetic (EM) and strong interactions vanish above $4\pi f$, so the relevant coefficients in the $\beta$-functions are computed where the contributions by the particles lighter than the dilaton are only taking into account. Hence, the sum over all the particles can be split into sums over light and heavy states, where the dividing scale is the dilaton mass.   Since the DM and the SM are embedded in the common sector of the (approximate) conformal invariance, the dilaton serves as a SM portal operator. 
The dilaton may fluctuate around $v =  246$ GeV, it also may decay into the photons, the leptonic pairs, dark photons, where the latter may be identified via the light hadronic and leptonic states or via the invisible $\bar\nu \nu$ pairs in the final states [21].
The particle asymmetry in the fermion and the boson dark sectors may exist similar to one of the baryon asymmetry in the Universe. Non-annihilating boson matter,   
the DM scalars, may be organised into gravitationally bound state, called the "boson star" (BS) [22-24]. 
The gravitational instability of a spatially uniform state of a relativistic scalar field has been discussed in [25]. The stability of boson stars built out of DM scalars due to repulsive self-interactions was analysed in [26]. The particle replenishment in the BS with increasing the mass and the radius up to some critical values continue until a black hole (BH) formation. 

We propose that the BS contains the massive scalar fields in asymptotically flat space-time. 
For this end, we introduce a novel approach to the Higgs-dilaton couplings that incorporates the dilaton operator $O(x)$ in the form of $N$ "garland" scalar fields. In the model, we show how to avoid the "cosmic crunch" or "the cosmic non stabilisation" emerging from the large and negative vacuum energy in the ground state of the broken CFT. We show:\\
- how EW symmetry breaking affects the hidden sector by breaking its conformal symmetry, and how the infra-red (IR) divergence is avoided by the IR mass;\\
- how the hidden sector can shift the quartic Higgs coupling $\lambda$ away from the SM.\\
For the Higgs mass modified by hidden sector all the Higgs dynamical characteristics (the widths etc.) are also modified.
There is also obtained the relation between the Higgs VEV and the VEV of the dilaton where the latter is saved even if $ v= 0$. No IR divergence we have in the model that allows to make several experimental predictions for LHC and FCC in the decays of the dilaton and dark photons into the pairs of leptons.  

{\it Classical model with two scalar fields. -}
We use a minimal model where both the scalar sector of the hidden world 
and  the SM Higgs come from the same ultraviolet (UV) dynamics.  Being the low-energy effective theory of a scale-invariant UV theory with new boson objects (the BSs), the scale-invariant violating terms with the Higgs fields are coupled to the conformal scalar fields.
For this end, we introduce the scalar field operator $ X(x)$ of the BS 
  \begin{equation}
\label{e1}
X(x) \in \left \{z(x); O(x), \bar O(x)\right \},
\end{equation}
where $z(x)$ is the set of the physical states operators including the Higgs boson field $h$.
The operators 
 \begin{equation}
\label{e2}
 O(x) = \sum^{N}_{k=1} c_{k}\,\varphi_{k} (x), \,\,\,\, \bar O(x) = \sum^{N^{\prime}}_{\nu =1} c_{\nu}\,\varphi_{\nu} (x)
\end{equation}
are the field "garlands" where  $\varphi_{k} (x) $  and $\varphi_{\nu} (x)$ in (\ref{e2}) run over $N$  and $N^{\prime}$ scalar fields, respectively, including, in particular, the dilaton fields in the conformal sector and the radion fields [27]; $c_{k}$ and $c_{\nu}$  are the coefficients. In AdS/CFT correspondence, the dilaton field is dual to the radion [28] in the model where all the SM fields are localised  at the IR brane in the five-dimensional anti-deSitter space background. We consider the Higgs boson portal coupling with the dilaton garland.
One can define the arbitrary commutation relations with regards to $O(x)$ and $\bar O(x)$ because we did not say definitely about the LD and the equations of motion. We admit the operator $\bar O(x)$ can even have the resembles the model containing the dipole field satisfying the equation of the 4th order, $(\Box + m_{\bar O}^{2})^{2} \bar O(x) = 0$ [29] with the massless case being more singular. It is also closely related to the model in a study of the Higgs [30] where the scalar "ghost" state has been formally found from which the present model is distinguished by the coupling of the garland $O(x)$ with the physical Higgs.  
Thus, we introduced a scale-invariant UV theory of the BS above the scale $\Lambda \simeq 4\pi v$, where the scale-symmetry is spontaneously broken at $\Lambda_{SB} >> \Lambda$. Below $\Lambda$, there are only scalars $O(x)$ and $\bar O(x)$ from (\ref{e2}), in addition to the SM fields. At energies $ \Lambda < E < \Lambda_{SB}$ the potential for $X(x)$ is $V_{X} = \Lambda^{2} X^{+}X + \lambda_{X} (X^{+} X)^{2}$, where $X(x)$ is defined by the dilatation current, $D_{\nu}(x)$, as $\langle 0\vert D_{\nu}(x)\vert X(p)\rangle = i\,f_{X}\,p_{\nu}\,e^{-ipx}$,  $f_{X}\sim \Lambda_{SB}$. The BS field may be coupled to gravity with action
$$S = \int d^{4} x \sqrt {-g} \left (\frac{{\it R}}{16\,\pi\,G} - \frac{1}{2}\,g^{\mu\nu}\,\partial_{\mu} X\,\partial_{\nu} X^{+} - V_{X}\right ), $$
where $ \it {R}$ is the Ricci scalar, $G$ is the gravitational constant.

The total Hilbert space $\mathcal {H}$ contains the subspace $\mathcal {H_{I}}$  of the states with the physical fields in $z(x)$, while the addition to it is the subspace $\mathcal {H_{II}}$ related to the hidden sector [31]. Then, for the states $\Phi$ which are projected from $\mathcal {H}$ to $\mathcal {H_{I}}$ by the operator $P$ one has 
\begin{equation}
\label{e3}
\Phi = P\Phi + (1-P) \Phi = h + \phi,
\end{equation}
where $h\in \mathcal {H_{I}}$, $\phi \in \mathcal {H_{II}}$, $ P^{+} = P$, $P^{2} = P$, ${\vert\vert \Phi\vert\vert }^{2} = {\vert\vert h\vert\vert }^{2} + {\vert\vert \phi\vert\vert }^{2}$, $\mathcal {H} = \mathcal {H_{I}} + \mathcal {H_{II}}$, ${\vert\vert h\vert\vert }^{2} > 0$. In the $S$-matrix approach to the field theory, all the fields in $X(x)$ (\ref{e1}) are considered in terms of the asymptotic states. 
 The transition probability from $h$   to the states $\phi$ in (\ref{e3}) is
   \begin{equation}
\label{e4}
\Pi \sim \frac{{\vert \langle h\, S\,\phi \rangle \vert }^{2}}{{\langle h \, h \rangle} {\langle \phi\,\phi\rangle}}, 
\end{equation}
where $S$-matrix transforms a physical state in a mixture of the latter and the states in the hidden sector. Hence, we assume $\langle \phi\,\phi\rangle > 0$, otherwise, it is not understandable to have a negative probability.  
To avoid the uncertainties in the calculations due to the hidden sector in the presence also the indefinite metric one can restrict the space of the final states to the space with the physical (positive) metric. To compensate these  related to the restrictions, one can assume that the initial state (in the hidden sector) to be in the space larger than the physical state space with the requirement that the metric in that physical space be also the physical one (positive). 
The garlands in (\ref{e2}) are characterised  by $N (N^{\prime})$ discrete  mass parameters $\Delta^{2}$ related  to the scalar fields $\varphi_{k} (\varphi_{\nu})$ in the approximate conformal theory. The conformal invariance in the BS is controlled by the small "garland tablet" $\sim \Delta^{2}/N(N^{\prime})$.
Using (\ref{e4}) one can expect, e.g.,  the production of the scalar fields (dilatons) through the decays of the Higgs boson where the study of the scalars produced can be done via the registration of the leptonic pairs.
The most stringent limits to date on the branching fraction of the Higgs boson to the scalar LLPs, subsequently decaying to $\tau^{+}\tau^{-}$ and light quarks were presented by the CMS at the LHC [32].

To develop the model further, we consider two scalar states $X$ and $X^{\star}$ with complex conjugate masses $\upsilon$ and $\upsilon^{\star}$, respectively. 
One can suggest the  field transformation
$$ X(x) = \frac{ \omega\,\phi(x) + i\,\varkappa\,h (x)}{\sqrt {2}},\,\,\, X^{\star}(x) = \frac{\omega\,\phi(x) - i\,\varkappa\,h (x)}{\sqrt{2}}, $$
where $\omega$ and $\varkappa$ are the constants.
The LD is 
$$L = \frac{1}{2i}\left [(\partial_{\mu} X)^{2} - (\partial_{\mu} X^{\star})^{2}\right ] + \frac{1}{2i} \left (\upsilon^{{\star}^{2}} X^{{\star}^{2}} - \upsilon^{2}\,X^{2}\right ), $$
 where in order to go to the case of real $\upsilon^{2}$ the scale transformations $\phi \rightarrow a\phi$, $h \rightarrow a^{-1} h$ are useful with $a$ being the positive constant. 
 The solution for $h (x)$  is 
\begin{equation}
\label{e13}
h(x) = \frac{\omega}{\varkappa}\frac{b^{2}}{m^{2}_{h\phi}} \left [\phi (x) + f x_{\mu}B^{\mu} (x)\right ]  + C(x),
\end{equation} 
where $Re\, \upsilon^{2} \equiv m_{h\phi}^{2}$, $Im(a^{2}\upsilon^{2}) \equiv b^{2}$, and the free (auxiliary) field $C(x)$ obeys the Eq. $(\Box + m^{2}_{h\phi} ) C(x) = 0$. Here, we followed the formalism of the non-linear realisation of conformal symmetry with linearisation of the homogeneous Lorentz group [33-36]. In  (\ref{e13}), the vector field $B_{\mu}(x)$ is associated with the generator of special conformal transformation, while $\phi (x)$ is the dilaton field related with $B_{\mu}(x)$ as  $B_{\mu}(x) = \partial_{\mu} \phi (x) / (2f)$. The field $B_{\mu} (x)$ does not in reality have the Goldstone nature, it is the DP field [21]. The Higgs field $h(x)$  in (\ref{e13}) is defined by the dilaton field $\phi (x)$ and distorted by the DP field $B_{\mu}$ up to the arbitrary field $C(x)$. 
It is easy to check that (\ref{e13}) is also given in the form suitable  for canonical quantisation
\begin{equation}
\label{e14}
h (x) = \int \left [h (k) + \frac{\omega}{\varkappa} \frac{b^{2}}{m^{2}_{h\phi}} \left (1 -\frac{1}{2} i\,k\,x \right ) \phi (k) \right ]e^{-ikx}\,\delta (k^{2} - m_{h\phi}^{2})\, d_{4} k. 
\end{equation} 
The second term in the solution (\ref{e13}), $h_{2}(x) \sim  x_{\mu}B^{\mu}(x) $,
is the dipole part of the $h$ - field obeying the  equation $(\Box + m_{h\phi}^{2})^{2} \,h_{2} (x) = 0$. In the Abelian dark photon model (ADPM) [21], the derivative of the dilaton field is the main part of the  DP field $B_{\mu}(x)$ solution. Neglecting the interactions with spin 1/2 DM charged under the new $U^{\prime}(1)$ gauge group, the solution for $B_{\mu}(x)$ looks like 
\begin{equation}
\label{e144}
B_{\mu}(x)\simeq \frac{1}{m_{DP}} \left (\frac{\ae}{m^{2}_{DP}}\Box -1\right )\partial_{\mu}\phi (x),
\end{equation} 
where $m_{DP}$ is the DP mass, $\ae $ is the real parameter in the LD of ADPM. The second term in (\ref{e144}) is nothing other but the inverse Higgs condition [37] which has been suggested to describe the independent fluctuations of the vacuum in conformal field theory. 

The commutation relations of the model are
$ [\phi (x), \phi (x^{\prime})] = 0,  [\phi (x), h (x^{\prime})] = iD(x - x^{\prime}; m^{2}_{h\phi}) $,
$ [h(x), h (x^{\prime})] = iD(x - x^{\prime}; m^{2}_{h\phi})$, 
where the standard permutation function $D (x)$ obeys the following equations   
$$ (\Box + m^{2}_{h\phi}) D(x) = 0, \,\,\,\,\,\, (\Box + m^{2}_{h\phi}) \frac{\partial}{\partial m^{2}_{h\phi}} D(x) = - D(x).$$ 
In the $k$ - momentum representation one has 
\begin{equation}
\label{e1444}
[\phi (k), h(k^{\prime})] \delta(k^{2} - m^{2}_{h\phi}) = \epsilon (k^{0}) \delta (k + k^{\prime}), 
\end{equation}
$ [\phi (k), \phi (k^{\prime})] = 0,  [h (k), h (k^{\prime})] = 0$.
The energy-momentum operator $P_{\mu}$ is defined by $ [P_{\mu}, \phi (x) ] = - 2 i f B_{\mu} (x), $ $ [P_{\mu}, h (x) ] = -  i \partial_{\mu} h(x),$ where
\begin{equation}
\label{e14444}
P_{\mu} = \int \epsilon (k^{0})\,k_{\mu} \left [ \phi (k) h(-k) + \left (\frac {\omega}{\varkappa}\right ) \left (\frac{b^{2}}{4 m^{2}_{h\phi}}\right )\phi (k)\phi (-k) \right ] \delta (k^{2} - m^{2}_{h\phi}) d^{4} k.
\end{equation}
The first term in (\ref{e14444}) is due to (\ref{e1444}), while the nature of the second term is related to the linear term in $x_{\mu}$ in (\ref{e14}). In the 4-momentum space one has 
\begin{equation}
\label{e144444}
 [P_{\mu}, \phi (k) ] = - k_{\mu}\,\phi (k), \,\,\, [P_{\mu}, h (k) ] = - k_{\mu} \left [ h(k) + \left (\frac {\omega}{\varkappa}\right ) \left (\frac{b^{2}}{4 m^{2}_{h\phi}}\right )\phi (k)\right ].
 \end{equation}
The vacuum is defined by the lowest eigenvector $P_{0}$ in (\ref{e144444}): $\theta(k^{0})\,\phi (k) \vert 0\rangle = 0,$ $ \theta(p^{0})\,h(p) \vert 0\rangle = 0,$ $k^{0} > 0, p^{0} > 0.$

One can suppose $h (x)$ is the part of some asymptotic pattern of a more complicated theory containing other fields corresponding to the physical metric. The massless case of the model with the hidden field sector is rather singular
and has been discussed in the $S$-matrix approach in [21]. 

{\it Effective theory of hidden sector.  -}
Let us consider the operator $ O(x)$  in (\ref{e2}) which contains 
$ N$ scalar fields $\varphi_{k}(x)$ with the masses 
$\mu^{2}_{k} = \mu^{2}_{0} + (\Delta^{2}/{N})\,k, \,\,\, (k= 1,2,..., N). $
The  mass interval $\Delta^{2}$ admits the approximate scale invariance breaking in the particular spectrum with the background square mass $\mu^{2}_{0}$. If $\mu^{2}_{0} \neq 0$ and $\Delta^{2}$ is small but finite, the BS is unstable and can decay, e.g., into the lepton pairs, the primary (direct) photons, DPs and light quarks. 
Actually, in the limit $N\rightarrow \infty$ there will be the possibility to recover an approximate scale invariant mass spectrum. 
We assume the dark sector is secluded from the visible sector, communicating only via the scalar fields portal. A spin-1/2 DM particle $\chi$ with the mass $m_{\chi}$ coupled to a scalar particle (the mediator) can give rise to a large annihilation cross section through the Sommerfeld-like enhancement (SE) [38].  
If we assume the mixture between the scalars in the SM and the hidden  sector then there is the possibility to introduce two mass eigenstates, the "light" and "heavy" scalars, interacting with DM field as
$\sim g^{j}\,O_{j} \,\bar\chi\,\chi$, where $j$ runs over "light" and "heavy" states $O_{light}$ and $O_{heavy}$, respectively. 
 Here, 
$ O_{light} (x)= \cos\alpha\,h_{SM}(x) + \sin\alpha\,O(x), $
$ O_{heavy} (x)= -\sin\alpha\,h_{SM}(x) + \cos\alpha\,O(x), $
where $\alpha$ is the mixing angle (free parameter), $h_{SM}$ is the SM excitation about 246 GeV, $g^{light} = (m_{\chi}/f)\sin\alpha$, $g^{heavy} = g^{light}(\alpha + \pi/2)$ [14]. The conformal compensator field $\sim f e^{O_{j}(x)/f}$ couples uniformly the SM and the DM.
For each $O_{j}$ with the mass $\mu_{j}$ the induced Yukawa potential is 
 \begin{equation}
\label{e355}
K(r) = - \frac {{(g^{j})}^{2}}{4\,\pi\,r}\, e ^{-\mu_{j}\,r}.
\end{equation} 
In terms of the ladder diagrams involving the multiple exchanges of the $O_{j}(x)$, the result (\ref{e355}) must be resumed. The SE can only arise if the attractive force scalar mass $\mu_{j} < [(g^{j})^{2}/(4\pi)] m_{\chi}$ (the Compton wavelength effect). 
We find the lower bound on $m_{\chi}$ which is about 3.9 TeV where $\sin\alpha$ is estimated from a global fit for combining all Higgs production channels and decay modes at the 95 \%  C.L. [14],  $\vert (v/f)\sin\alpha \vert \sim O(10^{-2}) $, and we set $\mu_{light} = m_{h}$ = 125 GeV with $m_{h}$ being the mass of the physical Higgs boson. 

The correlation function  of the free field operator $O(x)$ is 
$$ \int d^{4}x \,e^{i\,p\,x} \langle 0\vert T\,O(x)\,O^{+}(0)\vert 0\rangle = \int^{\infty}_{0} \frac{d t}{2\,\pi}\,\frac{\rho (t, \Delta^{2}; N)}{p^{2} - t + i\epsilon}, $$
where  the spectral density $\rho (t, \Delta^{2}; N) =  2\pi\,\sum_{k=1}^{N} {\vert c_{k}(\mu_{k}^{2}, \Delta^{2};N)\vert}^{2}\,\delta (t - \mu^{2}_{k})$. On the other hand, having in mind the scaling dimension $d$ of the operator $ O(x)$,  the $\rho (t)$ can be chosen as $\rho(t) = a_{d}\,t^{d-2}$ in the scale invariant case, where the propagator $\sim a_{d} (p^{2} + i \varepsilon)^{d-2}$ with $1< d < 2$. The precise form of the normalisation constant $a_{d}$ dependent on $d$ is not important for further studies. One can easily find 
$$ {\vert c_{k}(\mu_{k}^{2}, \Delta^{2};N)\vert}^{2}  = \frac{a_{d}}{2\pi}\,\frac{\Delta^{2}}{N} (\mu^{2}_{k})^{d-2}  $$
which is the degree of scale invariance breaking in the BS controlled by the very small "garland tablet" $\Delta^{2}/N$ with $\mu^{2}_{0} = 0$. Each $k$th dilaton particle "layer" is coupled with a strength $\sim  c^{2}_{k}$. In the pure conformal sector the dilaton can never decay ($\Delta = 0$, $\mu^{2}_{0}\rightarrow 0$). In the approximate conformal theory $\Delta \neq 0$ and the lifetime of $\varphi_{k}$ is $\sim (N/\Delta^{2})(\mu^{2}_{k})^{2-d}$.
As a consequence to that, SE is significant in the limit $\mu_{k}^{2}\rightarrow 0$. It means the DM can not escape the Yukawa domain (potential) well while the Yukawa forces are strong enough and the range of the forces is long enough
to contain DM. Clearly, the SE through the scalar exchange contribution $K_{j}(r)$  (\ref{e355}) becomes significant, saturates as $\sim m_{\chi}/f$ and  is most important for heavy DM and the light scalar,
 $\mu_{j}\langle r\rangle < O(1)$. 
The non-zero mass of the dilaton, $\sim f$, prevents the ultimate overproduction the photons  and dark photons  originated from the decays of the finite numbers of dilatons. The more details of the SE in the scattering cross section for the scalar can be found in [9].  The dilaton mass is a cut-off in the Yukawa potential to explain the modification of the high energy interactions. The scalar field $O(x)$ may also establish into gluon-gluon bound states having the vacuum quantum numbers $J^{PC} = O^{++}$, "the glueballs", of the pure Yang-Mills hidden sector in the early Universe. The lightest glueball has a mass estimated to be $\sim 10 T_{c}$ [39], where $T_{c}\sim 160$ MeV is the typical temperature in the SM of strong interactions. 

A systematic way to study the hidden scalar sector is to take advantage of the (approximate) scale invariance in order to build an effective Lagrangian for energies below $\sim 4\pi f$ where scale invariance is preserved by means of the dilaton field with the presence of the Higgs field.
We shall concentrate on the scalar potential only, as other interactions are not relevant for our analysis. The scalar potential is
 $W (H, \varphi_{k}) = \sum_{j=1}^{2} W_{j} (H, \varphi_{k})$, where the minimal scalar portal operates with the potentials
$$ W_{1} = - m^{2}H^{+}H + \lambda {\left (H^{+}H - \alpha^{2}\sum_{k=1}^{N} \varphi^{2}_{k} \right )}^{2},$$
 \begin{equation}
\label{e288}
 W_{2} = \xi\frac{m^{2}}{\Lambda} H^{+}H\, O + \lambda_{3}\,O\,\sum_{k=1}^{N} \varphi^{2}_{k} + \frac{1}{4}\zeta {\left ( \sum_{k=1}^{N} \varphi^{2}_{k} - f^{2}\right )}^{2}.
 \end{equation}
 Here, $m^{2}$ and  $\lambda$  are the mass parameter and  the coupling relevant to the Higgs $H$, respectively; $\xi$ and $\lambda_{3}$ are the $d$ - dependent couplings with the scale dimensions $(1 - d)$ and  $(2 - d)$, respectively; $\Lambda$ is the scale at which the theory becomes strong coupled. For large enough couplings $\lambda$ the fluctuations of $H$ about its VEV $\langle H^{+}H\rangle = \alpha^{2}\langle \sum_{k=1}^{N} \varphi^{2}_{k}\rangle$ are ultimate high, and can be integrated out. 
 The potential $W(H,\varphi_{k})$  is invariant under the $\mathbb {Z}_{2}$ background symmetry (BGS), where the latter  is explicitly broken by $\lambda_{3}$ and $\xi$. 
 All the observables (the measurable quantities) must be invariant under BGS. 
 The model may bring two effects: the mixing (coupling) between $H$ and $ O$ and the (invisible) decay, in particular, the decay of the scalar $O$ if $\Delta^{2}$ and $N$ are both finite. In the case the symmetry is unbroken, the scalar singlets $ O$ are stable and weakly interacting. Once the Higgs field accepts the VEV $v$, the term $\xi (m^{2}/\Lambda) {\vert H\vert }^{2}\, O$ in (\ref{e288})  brings the mixing $\sim v\,h\,O$ and the conformal scale is breaking because of the tadpole term $\sim h^{2}\,O$, where the EW doublet of the Higgs boson field is $H = (v + h)/\sqrt{2}$. The term $\sim \zeta f^{2}\,\sum_{k=1}^{N} \varphi^{2}_{k}$ in $W_{2}$ breaks the conformal invariance explicitly, and to make that breaking small we assume that $\zeta < 1$. In the limit  $\zeta\rightarrow 0$ the field associated with $\sum_{k=1}^{N} \varphi^{2}_{k}$ becomes a flat direction identified as a dilaton.
 In the Higgs field configuration $\langle H^{0}\rangle = h/\sqrt{2}$ the potential is $W(h,\varphi_{k}) = W_{1}(h) + W_{2}(h,\varphi_{k})$, where $W_{1} (h) = -(1/2)m^{2}\, h^{2} + (1/4)\lambda\, h^{2}$,
 \begin{equation} 
\label{e200}
\begin{gathered}
W_{2}(h,\varphi_{k}) =  -\frac{1}{2}\kappa \sum_{k=1}^{\infty} \mu^{2}_{k}\varphi_{k}^{2} + \lambda_{3} \left (\sum_{n=1}^{\infty} \varphi_{n}^{2}\right ) \sum_{k=1}^{\infty} c_{k}\varphi_{k}  +  \frac{1}{4}\lambda_{4}\left ( \sum_{k=1}^{\infty} \varphi_{k}^{2}\right )^{2}  \\
 +  \xi\,\frac{m^{2}}{2\Lambda}\,h^{2}\sum_{k=1}^{\infty} c_{k}\varphi_{k} - \lambda_{h\varphi}\,h^{2}\sum_{k=1}^{\infty} \varphi_{k}^{2} + \frac{1}{4}\zeta\,f^{4},
\end{gathered}
\end{equation} 
where we have used the ansatz $\sum_{n=1}^{N} \varphi_{n}^{2} f^{2} \rightarrow \eta \sum_{n=1}^{\infty}\mu^{2}_{n}\, \varphi_{n}^{2}$, $0 < \eta < \infty$ is the parameter that characterises the deviation from an exact scale invariance;
$\lambda_{4} = \zeta+ 4\,\lambda_{h\varphi}\alpha^{2}$, $\lambda_{h\varphi} = \lambda\alpha^{2}$, $\kappa = \zeta\eta$ is the flat direction degree. After EW symmetry breaking the singlet field $\varphi_{k}(x)$ will mix with the SM Higgs that can significantly change the Higgs boson properties search. 
The coupling $\lambda_{h\varphi}$ in (\ref{e200}) should not be small in such a way that the dilaton is in thermal equilibrium with the Higgs at high temperatures $T$. As the latter drops down to $T\sim \mu$, the heaviest dilaton modes with the mass $\mu$ freeze out and decouple from the thermal bath. After freeze-out the scalar garland can decay into lepton pairs, the photons (including the dark photons $\bar\gamma$), the DM and neutrinos. The $\bar\gamma $ would later decay to leptons, $\bar \gamma\rightarrow \bar l l$, and we are in position to have a relic density of $\bar\gamma s$ which is to be compared with the spin-1 DM relic density found in cosmological observations.  
The lightest scalar glueball states may lead to very massive and dense boson stars [40,41,26] if the self-interactions $\sim\lambda_{3}$, $\sim\lambda_{4}$ are repulsive.

The coupling $\lambda_{h\varphi}$ can lead to pair-production of the dilaton but can not induce its decay. 
It is assumed the large $\lambda_{h\varphi}$ will dominate the production of two dilatons, $\phi\phi$, e.g., in the decays of the Higgs boson, $h\rightarrow \phi\phi$.  The production rate of the dilatons is given by the probability 
$$ \Gamma (h\rightarrow \phi\phi) = \frac{\lambda^{2}_{h\varphi} v^{2}}{{8\pi} p^{0}_{h}} \sqrt { 1 - \frac{4\mu^{2}}{m^{2}_{h}}},$$
where $p^{0}_{h}$ is the energy of the physical Higgs boson. The two scalar fields production with fusion of $W^{+}W^{-}$, $ZZ$, $f\bar f$ (including the spin 1/2 DM) and $hh$ dependent also on $\lambda_{h\varphi}$ are suppressed by the c.m.s energy of the initial particles at their high energies.  The operator $O$ is associated with the scalar LLPs with the mass below $\sim$ 60 GeV that is the 
the lower bound of the dilaton mass [17], and each of the scalar can decay into the pairs of leptons and/or light quarks (see, e.g., [32]).  
One can assume the branching ratio $Br(h\rightarrow \phi\phi) \sim 10^{-2}$ in order to be complimentary to the LHC searches for the Higgs to invisible channels (see, e.g., [19]).

The minimisation equation for $\varphi_{k}$ is affected at $\langle \varphi_{k}\rangle = f_{k}$ and ${\langle H\rangle}^{2} = v^{2}/2$
 \begin{equation}
\label{e29}
f_{k} =\frac{ \xi\,(m^{2}/\Lambda)\,v^{2} + \lambda_{3}\,\sum_{n=1}^{\infty} f^{2}_{n}}{2\lambda_{h\varphi}\,v^{2} -  \lambda_{4}\sum_{m=1}^{\infty} f^{2}_{m}  + \kappa \mu^{2}_{k}}\cdot c_{k},
\end{equation} 
where the result is non-zero and is provided by the VEV of the Higgs.
The approximate scale invariance can still be used to understand the low-energy dynamics where in the continuum limit we have the result away from the limits $\lambda\rightarrow\infty$ and $\zeta\rightarrow 0$:
 \begin{equation}
\label{e30}
\langle O\rangle = \frac{a_{d}}{2\pi}\, \kappa^{1-d}\,\frac{ \xi\,(m^{2}/\Lambda)\,v^{2} + \lambda_{3}\,\sum_{n=1}^{\infty} f^{2}_{n}}{[2\lambda_{h\varphi}\,v^{2}  - \lambda_{4}\sum_{m=1}^{\infty} f^{2}_{m}]^{2-d}}\,\Gamma(d-1)\Gamma(2-d)
\end{equation} 
which is finite for $1 < d < 2$. Note that the propagator of $ O$ is $\sim (p^{2} +i\epsilon)^{d-2}$ and is also valid at  $1 < d < 2$.
Taking into account the squaring of (\ref {e29}) $f^{2} = \sum_{n = 1}^{\infty} f^{2}_{n}$ the upper limit on $f^{2} $ is 
\begin{equation}
\label{e321}
f^{2}  < 2\,v^{2}\,\frac{\lambda_{h\varphi}}{\lambda_{4}}.
\end{equation}
The limit (\ref{e321})  decreases when $\zeta$ goes away from the flat limit direction $\zeta\rightarrow 0$.
In the case the scalar dilaton field becomes a flat direction one has $ \langle  O\rangle\rightarrow\infty$ (a dilaton condensate).
  The Eq. (\ref{e30}) can be solved numerically. Actually, $ \langle O\rangle \neq 0$ even if $v^{2} =0$. The IR divergence is avoided by the IR mass 
$M^{2}_{IR} = 2 \lambda_{h\varphi} (v^{2} - 2\alpha^{2} f^{2}) - \zeta f^{2}$ under the limit (\ref{e321}),
while the term $\sim\lambda_{3}$ in the potential $W$ breaks the association of the finite $\langle  O\rangle $ with the EW symmetry breaking (EWSB). 
Under the assumption of a coupling structure similar to that of the SM, the physical Higgs boson width is constrained to be $3.2^{+ 2.8}_ {- 2.2}$ MeV while the expected constraint based on simulations is $4.1^{+ 5.0}_{- 4.0}$ MeV [42]. For the lower bound on the dilaton mass, $\mu = 60$ GeV, the Higgs - dilaton coupling $\lambda_{h\varphi}\simeq 0.03$ at the maximal energy of the Higgs boson production at the LHC. 
Then, the value of the IR mass is restricted by $M_{IR} < $ 60 GeV.  
Note, that in the paper [43], the dominant contribution to the singlet fermion DM $\chi$ relic abundance comes from the decay of the dilaton after freeze-out when $\lambda_{h\varphi}\simeq 0.01$. On the other hand, the freeze-in mechanism gives a major contribution when $\lambda_{h\varphi} > 0.02$. The small $\lambda_{h\varphi}$   leads to a large relic density of the dilaton after freeze-out which then gives larger contribution to $\chi$ relic density after dilaton decay.
 At this stage, one can estimate an upper limit to the mass of the BS,  $ M^{star}_{max}$, based on the formalism of self-interacting scalar fields [40]. 
 At the center of the cold BS the scalar particles can occupy the lowest energy level of the momentum $p\sim \pi\,R^{-1}$ (the Heisenberg principle of the uncertainty), where $R$ is the radius of the BS. In the relativistic case, $p\sim \mu$, and the BS mass $M^{star} \sim R/G = M^{2}_{Pl}/\mu$, where $M_{Pl} = 1/\sqrt {G}  \simeq 1.2\times 10^{19}$ GeV is the Planck mass. The characteristic energy density inside the system with very weak self-coupling $\lambda_{4}\rightarrow 0$ is $\rho_{E}\sim M^{2}_{Pl}\,\mu^{2}$. On the other hand, $\rho_{E} \sim \zeta \sum_{ k}\varphi_{k}^{2}\,f^{2}$. At finite $\lambda_{4}$, the BS configuration of gravitational equilibrium may be parametrised by
 $$ \digamma^{eq} = \frac{\lambda_{4}\,M_{Pl}^{2}}{4\pi\,\zeta\,f^{2}}.$$
 At energies $\sim 4\pi f$, the maximal mass of the star  scales as $M_{max}^{star} \rightarrow \sqrt {\digamma^{eq}}\,M^{star}$ with $M_{max}^{star}$ being
\begin{equation}
\label{e311}
 M^{star}_{max} = \sqrt \frac{{\lambda_{h\varphi}}}{2\pi\,\zeta} \, \frac{v\,m_{p}^{2}}{\mu\,f^{2}}\,M_{wd}\simeq 0.32\cdot 10^{-7}\zeta^{-1/2}\,\,M_{wd}
 \end{equation}
compared to the Chandrasekhar  mass of the white dwarf $M_{wd} \sim M_{Pl}^{3}/m_{p}^{2}$,  
$m_{p}$ is the proton mass and we used $\lambda_{h\varphi}\simeq 0.03$ with the light dilaton mass $\mu \sim m_{h}/2$. 
There is unbounded value for $M^{star}$ in the exact flat space-time: the maximal value (\ref{e311}) is conjugate to the finite flat direction degree $\zeta$. 
The maximal radius of the BS is of the order of the Schwarzschild radius, $R_{max}^{star} \sim R_{Sch}$, in particular, 
$$ R_{max}^{star} = G M^{star}_{max} = \sqrt \frac{{\lambda_{h\varphi}}}{2\pi\,\zeta} \, \frac{v}{f}\,\frac{M_{Pl}}{\mu\,f}.$$
If the BS does not explode, the star configuration at $\zeta\rightarrow 0$ can collapse to a BH with a very large radius.
To avoid this, we suppose $\sqrt{\zeta}\,f \sim \mu$ and can immediately get $  M^{star}_{max}  \simeq 1.7\cdot 10^{-6}\,M_{wd}$, while $ R_{max}^{star} \simeq 1.8\cdot 10^{11} GeV^{-1}$. 
Concerning the BS existence longevity in time, let us note it is governed by the principles of symmetry.
In particular, the lifetime of the BS depends on the lifetime of the scalar garland $O$ mixed with $\vert H\vert^{2}$ in (\ref{e288}). 
In the approximate $\mathbb {Z}_{2}$ symmetry, the operator $O$ is the LLP if $\xi << \Lambda\,v/m^{2}$.
The lifetime of $O$, $\tau_{O}$, is defined by the lifetime $\tau_{h}$ of the physical Higgs up to mass scale factor $\sim \xi\,v/\Lambda$. The latter allows the Higgs can be integrated out in (\ref{e288}), $m^{2}\vert H\vert^{2}\rightarrow O^{(h)}_{SM}$ with $ (a_{(h)}/v)\, O^{(h)}_{SM}(x)\, O(x)$, where $a_{(h)} = (\xi\,m^{2}\,v)/(\Lambda\,m^{2}_{h})$. The result for the $O$ - scalar lifetime is  $\tau_{o} = a_{(h)}^{2}\,\tau_{h}$ at $\mu \simeq m_{h}$. If we assume the BS is a long-lived state composed of long-lived scalar garlands with the lifetime $\tau_{o} > 1.3\cdot 10^{-7} s$, one has to require the small value of the parameter $\xi < 4\cdot 10^{-2} << \Lambda\,v/m^{2}$ at $\Lambda \sim O(M_{NP})$. Here, we have used the data on the branching fraction of the Higgs to two LLPs decaying to $\tau^{+}\tau^{-}$ with the proper decay length $c\tau > 40\, m$ and the LLP mass of 40 GeV [32]. The lifetime of the physical Higgs $7.7\cdot10^{-23}  s <\tau_{h} < 1.3\cdot 10^{-21} s$  is used based on the measurement of the Higgs decay width $\Gamma_{h} = 3.2^{+2.4}_ {- 1.7}$ MeV [44].

Let us examine the impact of the scalar hidden sector effects on EWSB. Looking through the Higgs basis one can find   
\begin{equation}
\label{e32}
\lambda v^{2} = m^{2} \left (1 - \frac {\xi}{\Lambda}\langle O\rangle \right )(1 + \delta), 
\end{equation} 
where $\delta = (\sqrt{2}\,\alpha f/v)^{2} <1$, $\langle O\rangle $ is given in (\ref{e30}) 
and $f^{2}$ can be solved numerically from the Eq.
\begin{equation}
\label{e322}
 f^{2} = \frac{a_{d}}{2\pi}\,{\left ( \frac{1}{\kappa}\right )}^{d-1}\, \frac{ [ \xi\,(m^{2}/\Lambda)\,v^{2} + \lambda_{3}\, f^{2}]^{2}}{[2\lambda_{h\varphi}\,v^{2}  - \lambda_{4}\, f^{2}]^{3-d}}\,\Gamma(d-1)\Gamma(3-d)
 \end{equation} 
at $ 1 < d < 3$. 
The expression (\ref{e32}) is the modification of the SM Higgs mass $m^{2}_{h} = 2\lambda v^{2}$ due to effects of hidden dilaton sector with the couplings $\sim [(\xi\,m^{2}/\Lambda)\, O - 2\lambda_{h\varphi}\, \sum_{k=1}^{N} \varphi^{2}_{k}] H^{+}H. $ 

{\it  Experimental constraints -}




A. In the early Universe at very high energies $E \sim O(M)$ with $M$ being the scale where the coupling constants run towards zero, the LD of the effective theory has the form
\begin{equation}
\label{e15}
L_{U}\sim M^{-k} \,O_{SM}\,O_{UV}, \,\,\, k = d_{UV}-4+n.
\end{equation} 
We follow from the standard assumption that both the SM and the hidden  sector are embedded in some strongly coupled sector and do not mix with elementary, weakly coupled fields.  Then  at energies $v < f < E < M$ the  conformal symmetry  is breaking down that may trigger the breaking of EW symmetry, and the LD (\ref{e15}) can get the form 
 \begin{equation}
\label{e16}
L_{U} \rightarrow L =  \epsilon_{k}\, O_{SM}\, O_{hidden},
\end{equation} 
where 
$\epsilon_{k} =  M^{-k}\,\Lambda^{d_{UV} - d}$ with $\Lambda < M$ being the  scale  at which the theory becomes strong coupled. Here, we assume that there exists a complete theory describing all the interactions in conformal sector which after integrating out all degrees of freedom above some large scale $M_{NP}$ fixes the "bare" action of the effective field theory. 
The dimension of the operator $O_{hidden}$ in the hidden sector is $d$. It is supposed that $\epsilon_{k}\rightarrow 0$ as $E\rightarrow f\sim v$. 
The hidden sector including the dilaton and the DP begins to show up as $E\sim \Lambda (\sim \epsilon_{k})$ increases. 
For the DP physics, $\epsilon_{k}$ may be interpreted in terms of the kinetic mixing angle $\varepsilon$ at $E\sim O(f)$. 
In $S$-matrix approach with the DP $\bar\gamma$ one has
$$ {\vert \langle S\,\bar\gamma_{out}\vert \varepsilon\, O_{hidden}\, O_{IR}\vert S\,\bar\gamma_{in}\rangle \vert }^{2} = \varepsilon^{2} \langle S\,\bar\gamma_{out}\vert O_{hidden}\vert S\,\bar\gamma_{in}\rangle \langle f\vert O_{IR} \vert 0\rangle, $$
where $O_{IR}$ is the operator in IR. 
The result of the DP discovery would be at the level $\epsilon_{k}^{2} \sim \varepsilon^{2}$ through either the leptonic final states (e.g., $e^{+}e^{-}$,  $\mu^{+}\mu^{-}$pairs) or in the missing energy and the momentum.
The latter channel means the decays $\bar\gamma\rightarrow\bar\nu\nu$ in antineutrino ($\bar\nu$) and neutrino ($\nu$) pair. 


The scale invariance of strong dynamics given by the LD (\ref{e16}) is spontaneously broken at the scale $f \geq v$. This dynamics is also responsible for breaking of EW symmetry at $v$. Below the scale $\Lambda$ the most important form of the operator $\sim O_{SM}\,O_{hidden}$ is 
\begin{equation}
\label{e17}
L_{hidden} \sim\bar\epsilon_{k}\, {\vert H\vert}^{2}\, O_{hidden} 
\end{equation} 
which leads to the nonzero values of $\langle  O\rangle$ (\ref{e30}) and $f^{2}$ (\ref{e322}). The scale factor in (\ref{e17}) is $\bar\epsilon_{k} = \Lambda^{d_{UV} - d}/M_{NP}^{d_{UV} -2} $.  
In this case, because  of the Higgs VEV in the amplitude of the process involving the hidden sector, it will cause the hidden stuff to move from its conformal fixed point.
 Below the scale $\bar\Lambda = \left (\bar\epsilon_{k}\,v^{2}\right )^{1/(4-d)}$ the theory becomes non-conformal.
The new physics will be visible through the observable $\hat\Delta$ (the measurable quantity)  when the energy $E$ of the experiment obeys the relation 
$\hat\Delta = \epsilon^{2}_{k}\, E^{2(d + n - 4)}$. The latter is the consequence of the couplings (\ref{e17}).
Using the constraint that the typical value of $E$ should exceed $\bar\Lambda$, one can estimate $\hat O$ for the configuration of the LD (\ref{e17})
$$\hat\Delta < {\left (\frac{E^{n}}{v^{2}\,M_{NP}^{n-2}}\right )}^{2}, $$
where no unphysical dimensions $d_{UV}$ and $d$ are involved in this bound. 

The different scenarios must be employed in order to extract the signal of NP, e.g., the dilaton and the DP, depending on precise values of the dilaton mass or the DP mass and the kinetic mixing with the ordinary photon. 
The possibility of enhanced Yukawa couplings of the $\varphi$-dilaton to leptons would imply new discovery channels that are not significant in the case of the SM Higgs [10]. Hence, the $\varphi \bar l l$ vertex would result in new discovery channel with the  signature $\varphi\rightarrow \bar l l$ which is negligible for the SM Higgs  due to the small Yukawa couplings.
In the low-energy case that is relevant to collider physics the dilaton couplings with leptons $l$ are 
\begin{equation}
\label{e177}
 L_{\varphi} = -\frac{\varphi}{f}\sum_{l} (m_{l} + \varrho\,y_{l}\,v) \bar l l,
 \end{equation}
where $l$:$e,\mu$; $\varrho = \mu^{2}/f^{2}$ parametrises the degree of the exact scale symmetry deviation; $y_{l}$ is the strength of the "anomalous" Yukawa coupling $\bar l\varphi l$; $m_{l}$ is the mass of the lepton. There is the presence of shifts in the dilaton Yukawa couplings to leptons in (\ref{e177}). The details of the symmetry breaking where the Yukawa couplings to $\varphi$ consist both of a term proportional to $m_{l}$ and a lepton mass independent term can be found in [10].
 From the experimental point of view, the signal of NP increases with energy and the effects of the hidden sector on observables for $\varphi\rightarrow \bar l l$ are bounded by 
 $$\hat\Delta_{\varphi \bar l l} < {\left ( \frac{E^{3}}{M^{2}_{NP}}\right )}^{2} {\left (\frac{m_{l} + \varrho\,y_{l}\,v}{v^{2}}\right )}^{2}, $$
 where  $M_{NP}  \sim 10^{5}$ TeV [19] and we used the gauge invariant operator $\sim \bar l \varphi l$ which leads to an effective $ n = 3$ operator $(f/M_{NP})\bar l l$. 
For di-muon channel, $gg\rightarrow \varphi\rightarrow \mu^{+}\mu^{-}$, one finds for the LHC (for illustration we set $y_{l} = 1$): $\hat\Delta^{LHC}_{\varphi\mu^{+}\mu^{-}} < (10^{-18} - 10^{-17})$ for dilaton mass range $60 - 160$ GeV, respectively, while the FCC case is the most promising window for the discovery, $\hat\Delta^{FCC}_{\varphi\mu^{+}\mu^{-}} < (10^{-11} - 10^{-10})$ with respect to the dilaton mass interval above mentioned. 
Assuming the current and future colliders can detect deviations from the SM of the order $\hat\Delta_{\varphi\mu^{+}\mu^{-}} \sim 1 \%$, one can see that the NP with di-muon production will be visible at the LHC as long as $M_{NP} < $ 20 TeV while the $M_{NP}$ for search at the FCC is restricted by $\sim$ 316 TeV. 
 The null search result for dilatons production and their decays into final leptonic pairs gives the limits on the model parameters, the dilaton mass $\mu$, the constant $f$, the strength $y_{l}$, the mass scale $M_{NP}$. 
New limits on $\mu$ and $f$ can be compared with those already obtained at the LHC.

The other decay channels are the Dalitz-decays of the dilatons,
$\varphi\rightarrow \gamma^{\star} \gamma, \bar\gamma\gamma$ with $\gamma^{\star}\rightarrow\bar l l$, $\bar \gamma \rightarrow l\bar l$ ($l: e,\mu$) where the effective interaction $\sim g\,B_{\mu}\,J_{em}^{\mu}$ 
implies the coupling of the DP field $B_{\mu}(x)$ with EM  current  $J_{\mu}^{em}$ with the strength $g = e\varepsilon$;  $e$ is the EM coupling and $\varepsilon$ is the kinetic mixing strength of 
the DP with the ordinary photon. In this case, the signal event rate of NP in the detector scales as $\sim \varepsilon^{2} \hat\Delta_{\varphi \bar l l}$, where $\varepsilon \sim 10^{-12} - 10^{-3}$ (see, e.g., the refs. in [21]). 
B. Finally, we consider the production of two di-muon pairs in decays of the dilaton, $\varphi\rightarrow\mu^{+}\mu^{-}\mu^{+}\mu^{-}$. The four charged lepton mode can provide the cleanest signature in terms of the peak in four lepton mass spectrum and small background (see, e.g., [45] for the case of the SM Higgs). In addition, the $\varphi\rightarrow four\, leptons$  decay mode allows to derive a precise mass measurement in different combinations of lepton final states. The total width of the dilaton and the mass are the subject of the search the double off-shell decays compared to that of the two photon decays of the dilaton. By the conformal invariance,
the only quarks lighter than the dilaton contribute in the quark loop to define 
the form-factor $F_{\varphi}(x_{1}, x_{2})$ 
for the process dilaton going to two virtual photons normalised to the form-factor $F_{\varphi} ( x_{1} = 0, x_{2} = 0)$ being the coupling constant where the dilaton going into two real photons. Here, $x_{i}$ $(i = 1, 2)$ is the sum squared of the four-momenta of the final $\mu^{+}$ and $\mu^{-}$ mesons. The main production of the dilaton at a hadron collider is due to gluon-gluon interaction, $gg\rightarrow \varphi$, similar to that of the minimal SM Higgs production mechanism. In this case, the significance of the dilaton signal is the significance of the Higgs signal times the combined the strong and the VEV coupling factor $\sim (c_{strong}^{\varphi})^{2}  (v/f)^{2}$. In the effective theory where the strong forces between the particles are bounded by the conformal sector of interactions, the coupling of the dilaton with gluons is $\sim c_{strong}^{\varphi} (G_{\mu\nu}^{a})^2\varphi/(8\pi f)$, where $G_{\mu\nu}^{a}$ is the standard gluon field strength tensor. The suppression $\sim (v/f)^{2}$ in the production cross section of the dilaton may be partly compensated by about tenfold squared increasing due to the dilaton coupling strength $c_{strong}^{\varphi} = (11 - 2 n_{light}/3)$  compared to $c_{strong}^{h} = 2/3$ for the case of the SM Higgs boson $h$ if the mass of the latter $m_{h} < 2 m_{t}$ ($m_{t}$ is the top-quark mass); $n_{light}$ is the number of quarks lighter than the dilaton in the effective loop of interactions. Thus, the large enhancement $\sim O(10^{2})$ in $\varphi g g$ coupling can indicate that the production channel $gg\rightarrow \varphi$ is very promising at hadron colliders. 

The variation of the signal cross section where the dilaton decays into two pairs of identical leptons $l$ with different charges  with respect to the actual Higgs signal in proton-proton ($pp)$ collisions is 
\begin{equation}
\label{e18}
k_{4l} = DR_{\varphi} (\gamma\gamma) \times \frac{R_{\varphi} (4l/\gamma\gamma)}{R_{h} (4l/\gamma\gamma)},
\end{equation}
where the detection ratio
\begin{equation}
\label{e19}
DR_{\varphi} (\gamma\gamma)  = \frac{\sigma (pp\rightarrow \varphi ...)}{\sigma (pp\rightarrow  h ...)}\cdot \frac{BR_{\varphi} (\gamma\gamma)}{BR_{h} (\gamma\gamma)}
\end{equation} 
reflects the detection event of the dilaton in $\gamma\gamma$ channel where $\sigma (pp\rightarrow s ...)$ is the production cross section of the scalar $s: h, \varphi$; $BR_{s}(\gamma\gamma)$ is the standard branching ratio for the $s$ decay into two photons. The cross section $\sigma$ in (\ref{e19}) is proportional  to the respective partial decay width to $gg$ in the approximation where the $gg\rightarrow s$ interaction is essentially point-like, and the QCD radiative corrections  to the $gg\rightarrow h$ and $gg\rightarrow \varphi$ subprocesses are nearly equal.  The ratio between partial decay widths of the dilaton to four leptons and two photons $R_{\varphi} (4l/\gamma\gamma) = \Gamma_{\varphi}(4l)/\Gamma_{\varphi}(\gamma\gamma)$ is 
\begin{equation}
\label{e20}
R_{\varphi}  \simeq \frac {\alpha}{(3\pi\mu^{3})^{2}}\int_{(2m_{l})^{2}}^{\mu^{2}} \frac {dx_{1}}{x_{1}} \int_{(2m_{l})^{2}}^{(\mu - \sqrt {x_{1}})^{2}}\frac {dx_{2}}{x_{2}} A(x_{i};m_{l},\mu)Y(\mu, m_{l};f; y_{l}){\left \vert \frac{F_{\varphi}(x_{1},x_{2})}{F_{\varphi}(0,0)}\right \vert }^{2},
\end{equation}
where 
$$ A = \lambda^{1/2} (\mu^{2};x_{1}, x_{2})  [(\mu^{2} - x_{1} - x_{2})^{2} + 2x_{1}x_{2}] \Pi_{i=1}^{2} \left [ \left (1 + \frac{2 m_{l}^{2}}{x_{i}}\right ) {\left (1 - \frac{4 m_{l}^{2}}{x_{i}}\right )}^{1/2}\right ], $$ 
$$ Y(\mu, m_{l};f; y_{l}) = {\left [ 1 + \frac{\mu^{2}\,y_{l}}{m_{l}\,v}\,{\left (\frac{v}{f} \right )}^{2} \right ]}^{2},$$
$\lambda (x,y,z) $ is the K\" al\' en triangle function.
The "crossed diagram" must be added to the result (\ref{e20}). Since the magnitude of each diagram is small, the interference between them will also give a small contribution. The ratio $R_{h}$ for the SM Higgs boson  [46] has the same form as that to $R_{\varphi}$ except the shifting factor $Y$. There is an enhancement of the EM $\gamma\gamma\varphi$ coupling (factor $\sim$ 4) for $\mu < 2m_{W}$ compared to the $h$ for the physical Higgs boson mass $m_{h}$ = 125 GeV. The quantity $[ 1/DR_{\varphi} (\gamma\gamma)]\times k_{4l}$ may be computed numerically for the appropriate form-factor squared ${\vert F_{s} (x_{1},x_{2}) \vert}^{2}$ ($s: h,\varphi$).  
The upper limit of $[ 1/DR_{\varphi} (\gamma\gamma)]\times k_{4\mu}$ for two pairs of muons in the final state is 
$$ [ 1/DR_{\varphi} (\gamma\gamma)]\times k_{4\mu} < {\left [\frac {\log(\mu/2 m_{\mu})}{\log(m_{h}/2m_{\mu})} \right ]}^{2}\times Y(\mu; f; y_{\mu}).$$
By use of $DR_{\varphi} (\gamma\gamma)$ in Fig. 3 in [12] for the dilaton mass $\mu$ = 100 GeV, $f$ = 3 TeV one has $k_{4\mu} <$ 0.9 where we set $y_{\mu} = 1$. The search result for $\varphi$ decaying into final states containing two muon pairs places limits on the model parameters $\mu$, $f$ and $y_{\mu}$. Because of the enhanced coupling $\varphi\mu^{+}\mu^{-}$, the rate $R_{\varphi} (4\mu/\gamma\gamma) $ may deviate significantly from that of the SM Higgs, $R_{h} (4\mu/\gamma\gamma) < 0.4\cdot 10^{-4}$ for $m_{h} = 125 $ GeV. The enhancement for the couplings $\varphi e^{+}e^{-}$ or $\varphi \tau^{+}\tau^{-}$ can also lead to other clean discovery channels, $\varphi\rightarrow  2e^{+} 2 e^{-}$ or $\varphi\rightarrow  2\tau^{+}2\tau^{-}$ for $\mu <$ 160 GeV. The only restrictions should be taking into account within the calculations: the electron mass should not go to zero to avoid the quadratic divergence leading to the violation of the Lee-Nauenberg theorem in quantum electrodynamics [47].


{\it  Conclusions -}
In the paper, we followed the statement  that the SM is embedded into quantum field theory with conformal symmetry where there are:\\
- combined conformal and EW symmetry breaking,\\
- the possibility to implications for beyond the SM phenomenology,\\
- the possibility to implications for Higgs couplings, the dilaton couplings, the DM and DP couplings. \\
We introduced the formalism to examine the impact of the hidden scalar sector effects on the conformal and the electroweak symmetries breaking. 
The novel approach to the scalar BS naturalness is considered. The BS is presented by the local scalar field containing the Higgs boson field and the garland-like scalar fields including the dilaton fields of the conformal field theory.  It also allows to stabilise the theory and to get the finite value of the VEV of the dilaton field.
 We show that taking into account the repulsive self-interactions ($\lambda_{4}$) and the flatness degree ($\zeta$) in the dark scalar sector prevents instability of the BS related to the BH formation. In addition, both $\lambda_{4}$ and $\zeta$ can provide the conditions to formation of the Bose-Einstein condensation.
The importance of the BS is actual and evident, because the boson stars hide a scalar part of DM from direct observation. After the formation and becoming large in size, the BS may explode into leptons via decays of dark photons or emit primary photons which may explain the conformal anomaly in EW and EM hidden sectors. These could contribute to new sources of cosmic rays.
We studied in details how the EWSB affects the hidden scalar sector by breaking its conformal symmetry and generating a mass gap to avoid the IR divergence.

 We applied our formalism to determine the modification of the self-coupling constant (\ref{e32}) of the Higgs boson away from its SM value within the influence of the hidden sector. 
 This leads to the changes in the properties of the Higgs, namely, its mass and its width, already at the tree level, making the Higgs and the dilaton a mixed sector.
 We have estimated the rate of the deviation from the SM with production of lepton pairs, e.g., the di-muons, due to decays of the dilaton and through the decays of the DP. The latter is  the subject of the dilaton decay first. 
 We make several predictions in the combined SM - hidden sector, in particular:\\ 
 - the DM mass, $m_{\chi} > $ 3.9 TeV;\\
 - the Higgs-dilaton coupling constant, $ \lambda _{h\varphi} \simeq 0.03$;\\
 - an upper limit of IR mass, $M_{IR} < 60$ GeV;\\
 - maximal mass of the boson star, $M_{max}^{star} \sim 1.7\cdot 10^{-6} M_{wd}$ ;\\
 - the NP scale in di-muon production with the  $\sim 1 \%$ deviation effect to the SM: $M_{NP}^{LHC} < 20$ TeV  and $ M_{NP}^{FCC} < 316$ TeV;\\
 - the variation in the $\varphi\rightarrow \mu^{+}\mu^{-}\mu^{+}\mu^{-}$ signal, $k_{4\mu} < 0.9$ (\ref{e18}).\\
 The confrontation of these bounds with the results of the collider experiments as well as the cosmological observation and the studies will constitute the further precise test of the boson star model.
 The results of the paper may be used in the experiments sensitive to NP with masses in TeV or multi-TeV range and strongly coupled to SM particles. 

\end{document}